\documentclass[10pt,a4paper]{article}

\usepackage{amssymb}
\usepackage{amsfonts}
\usepackage{mathrsfs}
\usepackage{latexsym}
\usepackage{amsmath}
\usepackage{amsthm}
\usepackage{amscd}               
\usepackage{graphics}
\usepackage{graphicx}
\usepackage{hhline}                   
\usepackage{euscript}               
\usepackage{rotating}
\usepackage{setspace}                 
\usepackage{fancyheadings}            
\usepackage[dvips]{hyperref}
\DeclareSymbolFont{NumBold}{U}{bbold}{m}{n}
\DeclareSymbolFontAlphabet{\mathNumbb}{NumBold}
\DeclareMathSymbol{\Id}{\mathord}{NumBold}{`1}

\newcommand{\D}{\mathcal{D}}
\newcommand{\E}{\mathcal{E}}
\newcommand{\F}{\mathcal{F}}
\newcommand{\G}{\mathcal{G}}

\renewcommand{\L}{\mathcal{L}}

\newcommand{\N}{\mathcal{N}}

\newcommand{\Q}{\mathcal{Q}}
\newcommand{\R}{\mathcal{R}}

\newcommand{\V}{\mathcal{V}}

\renewcommand{\a}{\alpha}
\renewcommand{\b}{\beta}
\newcommand{\g}{\gamma}
\renewcommand{\d}{\delta}
\newcommand{\eps}{\varepsilon}
\newcommand{\ep}{\epsilon}
\renewcommand{\o}{\omega}

\newcommand{\vf}{\varphi}
\renewcommand{\l}{\lambda}
\newcommand{\s}{\sigma}

\renewcommand{\th}{\theta}

\newcommand{\z}{\zeta}

\newcommand{\gG}{\Gamma}

\newcommand{\gL}{\Lambda}
\newcommand{\gTh}{\Theta}
\newcommand{\gO}{\Omega}
\newcommand{\gS}{\Sigma}

\newcommand{\bA}{\mathbb{A}}

\newcommand{\e}{\mathop{\mathrm{e}}\nolimits}

\newcommand{\Tr}{\mathop{\mathrm{Tr}}\nolimits}

\newcommand{\vpint}{\displaystyle -\hskip -1.1em \int}

\newcommand{\diag}{\mathop{\mathrm{diag}}\nolimits}
\newcommand{\Ch}{\mathop{\mathrm{Ch}}\nolimits}
\newcommand{\Td}{\mathop{\mathrm{Td}}\nolimits}
\newcommand{\Ind}{\mathrm{Ind}}

\newcommand{\Spin}{\mathop{\mathrm{Spin}}\nolimits}
\newcommand{\dd}{\mathrm{d}}
\newcommand{\pd}{\partial}
\newcommand{\cD}{\nabla}
\newcommand{\Real}{\mathbb{R}}
\newcommand{\Compl}{\mathbb{C}}
\newcommand{\Integer}{\mathbb{Z}}

\newcommand{\Sphere}{\mathbb{S}}

\newcommand{\Quat}{\mathbb{H}}
\newcommand{\Oct}{\mathbb{O}}
\let\IM=\Im
\let\RE=\Re
\let\Im=\undefined
\let\Re=\undefined
\newcommand{\Im}{\mathop{\IM\mathfrak{m}}\nolimits}
\newcommand{\Re}{\mathop{\RE\mathfrak{e}}\nolimits}

\newcommand{\Tor}{\mathbb{T}}

\newcommand{\tld}{\widetilde}

\newcommand{\<}{\langle}
\renewcommand{\>}{\rangle}

\newcommand{\w}{\wedge}

\newcommand{\SU}{\mathop{\mathrm{SU}}\nolimits}
\newcommand{\SO}{\mathop{\mathrm{SO}}\nolimits}
\newcommand{\Sp}{\mathop{\mathrm{Sp}}\nolimits}
\newcommand{\UU}{\mathop{\mathrm{U}}\nolimits}
\newtheorem{remark}{Remark}

\newcommand{\microsection}[1]{\noindent\underline{\textbf{#1}}}
\newcommand{\Ref}[1]{(\ref{#1})}
\newcommand{\blanc}{\ifthenelse{\boolean{@twoside}}{\thispagestyle{empty}~\newpage}{} }

\begin{document}


\thispagestyle{empty}

\makeatletter
\begin{titlepage}
\begin{flushright}
{hep-th/0611278}\\
\end{flushright}
\vskip 1cm
\begin{center}
{\LARGE\bf On F--term contribution to effective action}
\end{center}
\vskip 2cm
\begin{center}
{\bf Sergey Shadchin}
\end{center}
\vskip 1cm
\centerline{\em INFN, Sezione di Padova \& Dipartimento di Fisica ``G. Galilei''}
\centerline{\em Universit\`a degli Studi di Padova, via F. Marzolo 8, Padova, 35131, ITALY}
\centerline{\tt email: serezha@pd.infn.it}
\vskip 3cm
\centerline{\sc Abstract}
\vskip 1cm
We apply equivariant integration technique, developed in the context of instanton counting, to two dimensional $\N=2$ supersymmetric Yang--Mills models. Twisted superpotential for $\UU(N)$ model is computed. Connections to the four dimensional case are discussed. Also we make some comments about the eight dimensional model which manifests similar features.
\bigskip
\end{titlepage}
\makeatother


\tableofcontents


\section{Introduction}


\subsection{Motivation}

The fundamental algebraic result, known as Hurwitz theorem, claims that there are only four division algebras: $\Real$, $\Compl$, $\Quat$ and $\Oct$. Their real dimensions are $1,2,4$ and $8$ respectively. Another known fact concerns minimal supersymmetric models. Consider the dynamical gauge field interacting with fermions in $d$ dimensional space--time. Denote $\cD_I = \pd_I + A_I$ the covariant derivative, $F_{IJ} = [\cD_I,\cD_J]$ the curvature, $I,J = 1,\dots,d$. The action
\begin{equation}
-\frac{1}{4}F_{IJ}F^{IJ} + i \bar{\psi} \gG^I \cD_I \psi,
\end{equation}
can be on--shell supersymmetric only if $d = 3$, 4, 6 or 10. Clearly these dimensions can be written as $d = 2 + \dim_\Real \bA$, where $\bA$ is a division algebra.  This is not a coincidence. These minimal models capture the features of corresponding division algebras.

This fact is being observed since long time in various contexts \cite{Octonions,DAlgSphereTwist,SUSYandDiv,ParticleAndDiv}. Let us mention also that extended $\N=2$ and $\N=4$ supersymmetry in four dimensions can be obtained by dimensional reduction of $\N=1$ supersymmetry from six and ten dimensions respectively. Close relations of $\N=1$ and $\N=2$ supersymmetry in four dimensions with complex numbers and quaternions was figured out in \cite{QuatN=2SUSY}. 

In the context of supersymmetric Yang--Mills models this connection was pointed out, in particular, in \cite{GeneralizedInst} in the context of the Witten index calculation. It was shown that these contributions are given by regularized volumes of certain K\"ahler, hyper--K\"ahler and octonionic quotients.

Another interesting property of minimal supersymmetric Yang--Mills models is related to the dimensional reduction. Namely if we compactify 2 of $d$ dimensions in such a model, we obtain $\N=2$ supersymmetric model in $d-2$ dimensions. We will  not consider $d=3$ minimal supersymmetric model, and focus on $d-2 = 2$, 4 and 8. Common property of these theories is that they contain the topological sector \cite{JoyceD=8,BaulieuD=4N=1,BaulieuBRSTTQFT,BaulieuSinger8DandOther,Gey1,TQFT,IntroToCohFT,WittSUSYYM4Man}. An essential condition which makes possible the topological twist in eight dimensions is that the holonomy group of the manifold is $\Spin(7)$. Otherwise it should be an eight dimensional Joyce manifolds. One can show that the path integral for the vacuum expectation of a topological observable gets localized onto the moduli space of so--called generalized instantons \cite{GeneralizedInst}, which can be described as follows. Pick a complex structure on a $d$--dimensional manifold and define the symplectic form $\o = \sum_{i=1}^{d/2} \dd z^i \w \dd \bar{z}^i$. Generalized instantons are solution for the following equations:
\begin{equation}
\begin{aligned}
\o^{\frac{d-2}{2}}\w F^{(1,1)} &= 0, & F^{(2,0)} = F^{(0,2)} &= 0.
\end{aligned}
\end{equation}
In two dimensions it is equivalent to $F = 0$. To make theory reacher one can add some matter hypermultiplets. In such a way one obtains two dimensional Bogomol'ny equations. In four dimensions the general condition can be rewritten as self--dual equation: $F = \star F$, whereas in eight dimensions it becomes generalized self--dual equation: $F_{IJ} = \frac{1}{2}\Phi_{IJKL}F^{KL}$, where $\Phi_{IJKL}$ is eight dimensional $\Spin(7)$--invariant self--dual Caley tensor.  

In four and eight dimensions these equations can be written in components as follows:
\begin{equation}
\label{genInst}
\begin{aligned}
d &= 4 & &: & F_{4i} &=  \frac{1}{2}\ep_{ijk}F^{jk} \\
d &= 8 & &: & F_{8A} &= \frac{1}{2}c_{ABC}F^{BC},
\end{aligned}
\end{equation}
where $\ep_{ijk}$ is three dimensional Levi--Civita tensor, which is, at the same time, the quaternionic structure constants, $i,j,k = 1,2,3$, and $c_{ABC}$ are octonionic structure constants, $A,B,C = 1,\dots,7$. We see again the traces of algebras $\Quat$ and $\Oct$. Let us, in that follows, refer $d = 2,4$ and $8$ theories as $\Compl$--, $\Quat$-- and $\Oct$--case respectively.

In four dimensions $\N=2$ supersymmetric Yang--Mills models (which is the $\Quat$--case in our classification) was studied extensively both from mathematical (Witten approach to Donaldson invariants) and physical (Seiberg--Witten theory for low--energy effective action) point of view. The moduli space of $\Quat$--instantons are given by finite dimensional ADHM construction \cite{ADHM,SelfDualSolution,InstAndRec}, which identify the moduli space of instantons with certain hyper--K\"ahler quotient. It allows to reduce a path integral for the vacuum expectation of an observable to a finite dimensional (and hence well--defined) integral. 

It was shown by Nekrasov in \cite{SWfromInst} how to compute in this theory the partition function, which is given by the vacuum expectation of ``1''. After certain deformation of the model the partition function can be identified with the equivariant Euler characteristics of the instanton moduli space. A nice property of the deformed model is that this very quantity determines the leading term of the effective low--energy action. It is given by the F--term and is due to instanton contributions. Neither antiinstantons nor mixed instantons--antiinstatons do not contribute to the F--term \cite{NikitaLosevTauBar}. This conclusion holds both for models with and without matter hypermultiplets.

Since $\Compl$-- and $\Oct$--cases stand in a line with $\Quat$--case, it is natural to ask if the same is true in two and eight dimensional theories. Otherwise if in these theories the F--term of the effective action can be computed as the Euler characteristics of the appropriate moduli space.  

The purpose of present paper is to provide a partial answer to this question. Namely, we study in details the $\Compl$--case and show that under certain assumptions the answer is positive. We compute corresponding F--term via localization approach and provide some explicit formulae. Through the paper we display the similarities and distinctions with the $\Quat$--case. 

Two dimensional $\N=2$ supersymmetric models are well studied in the literature \cite{Freckled,Phases}. Less is known about the $\Oct$--case. Another purpose of this paper is to provide a basis for the intuition about $\Oct$--case, which will be object of the future investigations.

The paper is organized as follows. In Section \ref{C-model} we describe the dimensional reduction and topological twist of super Yang--Mills in four dimensions. Section \ref{Coh} is devoted to some cohomological aspects of the model. The path integral for topological observables localizes onto the vortex moduli space, which is described in Section \ref{vortex}. Section \ref{OmegaBack} is devoted to the $\gO$--background. Finally in Section \ref{Super} we compute the twisted superpotential.


\subsection{Notations and conventions}

Following notations are used through the paper: 
\begin{itemize}
\item The roman indices $I,J,\dots$ run over $0,1,2,3$. The greek indices (dotted and undotted) $\a,\b,\dot{\a},\dot{\b},\dots$ run over $1,2$, this is spinor indices. The 2--dimensional Lorentz indices are denoted by greek letters $\mu,\nu,\dots$ and run over $1,2$.
\item The generators of the Lorentz group are chosen as follows:
\begin{equation}
\begin{aligned}
\s^I_{\a\dot{\a}} &= {(\Id_2,-\tau_1,-\tau_2,-\tau_3)}_{\a\dot{\a}}, & \bar{\s}^{I,\dot{\a}\a} &= {(\Id_2,+\tau_1,+\tau_2,+\tau_3)}^{\dot{\a}\a},
\end{aligned}
\end{equation}
where $\tau_i$, $i=1,2,3$ are Pauli matrices in the standard form.
\item The 't Hooft projectors are defined as usual:
\begin{equation}
\begin{aligned}
\s^{IJ} &= \frac{1}{4} \left( \s^I \bar{\s}^J - \s^J\bar{\s}^I\right), & \bar{\s}^{IJ} &= \frac{1}{4} \left( \bar{\s}^I \s^J - \bar{\s}^J \s^I \right)
\end{aligned}
\end{equation}
\item The Grassman measure is defined in such a way that $\int \dd^2 \th (\th\th) = +1$, and the same for $\bar{\th}$. The twisted Grassman measure is defined as follows: $\dd^2 \vec{\th} = \dd \th_2 \dd \th_1$. It satisfies $\int \dd^2 \vec{\th} (\th_1\th_2) = +1$.
\end{itemize}

\bigskip
\microsection{Acknowledgments.} I thank Gianguido Dell'Agata, Marco Matone, Paolo Pasti and Roberto Volpato for their constant interest to this work. I am grateful to  Vladimir Fateev, Alexei Gorinov, Vladimir Lyakhovsky, Andr\'e Neveu and Mittier Pronob for helpful discussions. 

This work was partially supported by the EU MRTN-CT-2004-005104 grant ``Forces Universe'' and by the MIUR contract no. 2003023852.
 

\section{The $\Compl$--model}
\label{C-model}

In this section we briefly recall some key ingredients of the dimensional reduction from four to two dimensions as well as of the topological twist of a Yang--Mills model in two dimensions. 


\subsection{Four dimensional supersymmetry}

We start with $\N=1$, $d=4$ super Yang--Mills model in the Minkowskian space. Supersymmetric action for the theory without matter is described by the gauge multiplet, which can be arranged in a real scalar or a chiral spinor  superfield:
\begin{equation}
\label{gaugeSUFI}
\begin{aligned}
V(x,\th,\bar{\th}) &= i \th \s^I \bar{\th} A_I + i (\th \th) \bar{\th} \bar{\psi} - i (\bar{\th}\bar{\th}) \th \psi + \frac{1}{2}(\th\th)(\bar{\th}\bar{\th}) D, \\
W_\a(y,\th) &= i \psi_\a + \th_\a D - \s^{IJ}{}_\a{}^\b \th_\b F_{IJ} + (\th\th) \s^I_{\a\dot{\a}}\cD_I \bar{\psi}^I.  
\end{aligned}
\end{equation}
In the last line all fields are functions of covariantly constant directions in the flat superspace $y^I = x^I + i \th \s^I \bar{\th}$. Let $\z^\a$ and $\bar{\z}^{\dot{\a}}$ be parameters of supersymmetry transformation. Denote the supersymmetry operator as $\d = \z^\a \Q_\a + \bar{\z}^{\dot{\a}}\bar{\Q}_{\dot{\a}} = \z \Q - \bar{\z}\bar{\Q}$. The algebra is given by
\begin{equation}
\label{gaugeSUSY}
\begin{aligned}
\d A_I &= \z^\a \s_{I,\a\dot{\a}} \bar{\psi}^{\dot{\a}} - \psi^\a \s_{I,\dot{\a}\a} \bar{\z}^{\dot{\a}}, \\
\d \bar{\psi}^{\dot{\a}} &= -i \bar{\s}^{IJ,\dot{\a}}{}_{\dot{\b}} \bar{\z}^{\dot{\b}} F_{IJ} -i \bar{\z}^{\dot{\a}}D, \\
\d \psi_\a &= - i\s^{IJ}{}_\a{}^\b \z_\b F_{IJ} + i \z_\a D, \\
\d D &= -\z^\a\s^I_{\a\dot{\a}}\cD_I \bar{\psi}^{\dot{\a}} - \cD_I \psi^\b \s^I_{\b\dot{\b}}\bar{\z}^{\dot{\b}}.
\end{aligned}
\end{equation}
The action for the pure Yang--Mills theory can be put to the following form:
\begin{equation}
\label{Sgauge}
S_{gauge} = - \frac{1}{8}\left(\int \dd^2 \th W^\a W_\a + \int \dd^2 \bar{\th} \bar{W}_{\dot{\a}}\bar{W}^{\dot{\a}} \right) = - \frac{1}{8}F_{IJ}F^{IJ} + \frac{i}{2}\psi \s^I \cD_I \bar{\psi} - \frac{D^2}{4}.
\end{equation}
We have omitted the space integral and the trace over the adjoint representation of the gauge group Lie algebra, as well as trace normalization factors, for the sake of brevity. Note also the we have chosen unusual normalization for the action, which differs by factor 2 against the traditional one. This is done to avoid some $2\sqrt{2}$'s in relevant formulae. 

If the gauge group contains an $\UU(1)$ factor, we can also add the Fayet--Iliopoulos term which can be written as follows:
\begin{equation}
\label{FI}
S_{FI} = -2r \int \dd^2 \th \dd^2 \bar{\th} V = - r D.
\end{equation}

To add matter one can take two hypermultiplets which can be put into scalar chiral superfields $Q(y,\th)$ and $\tilde{Q}(y,\th)$. They are acted on by the gauge group in dual representations, in such a way that $\tilde{Q}Q$ be gauge invariant. The component expansion of a hypermultiplet is the following:
\begin{equation}
Q(y,\th) = q + \sqrt{2}\th \mu + (\th \th) f,
\end{equation}
where all fields are functions of $y^I$, and the same for $\tilde{Q}$. The supersymmetry acts as follows:
\begin{equation}
\label{matterSUSY}
\begin{aligned}
\d q &= \sqrt{2}\z^\a\mu_\a & \d \bar{q} &= \sqrt{2}\bar{\z}_{\dot{\a}} \bar{\mu}^{\dot{\a}}\\
\d \mu_\a &= i \sqrt{2}\s^I_{\a\dot{\a}} \bar{\z}^{\dot{\a}} \cD_I q + \sqrt{2}\z_\a f & \d \bar{\mu}_{\dot{\a}}&= -i\sqrt{2}\z^\a \s^I_{\a\dot{\a}} \cD_I \bar{q} + \sqrt{2}\bar{\z}_{\dot{\a}}\bar{f}  \\
\d f &= - i \sqrt{2}\cD_I \mu^\a \s^I_{\a\dot{\a}}\bar{\z}^{\dot{\a}} + 2i \bar{\z}_{\dot{\a}}\bar{\psi}^{\dot{\a}}q & \d \bar{f} &= i \sqrt{2}\z^\a \s^I_{\a\dot{\a}}\cD_I \bar{\mu}^{\dot{\a}} + 2i \z^\a \psi_\a \bar{q}.
\end{aligned}
\end{equation}
The action is given by 
\begin{equation}
\begin{aligned}
S_{matter} &= \frac{1}{2} \int\dd^2 \th \dd^2 \bar{\th} \left(Q^\dag \e^{2V} Q + \tilde{Q}^\dag \e^{-2V} \tilde{Q} \right) \\
&= - \frac{1}{2}\cD_I \bar{q} \cD^I q - \frac{i}{2} \mu \s^I \cD_I \bar{\mu} + \frac{i}{\sqrt{2}}\bar{q}\psi \mu - \frac{i}{\sqrt{2}}\bar{\mu}\bar{\psi}q + \frac{1}{2}\bar{q}Dq + \frac{1}{2}\bar{f}f \\
&\phantom{=} - \frac{1}{2}\cD_I \bar{\tilde{q}} \cD^I \tilde{q} - \frac{i}{2} \tilde{\mu} \s^I \cD_I \bar{\tilde{\mu}} - \frac{i}{\sqrt{2}}\bar{\tilde{q}}\psi  \tilde{\mu} + \frac{i}{\sqrt{2}}\bar{\tilde{\mu}}\bar{\psi}\tilde{q} - \frac{1}{2}\bar{\tilde{q}}D\tilde{q} + \frac{1}{2}\bar{\tilde{f}}\tilde{f}.
\end{aligned}
\end{equation}

The second (tilded) multiplet is needed only to introduce a supersymmetric mass in four dimensions. Corresponding contribution to the action is 
\begin{equation}
\label{4dMass}
\begin{aligned}
S_{MASS} &= \Re  \left(M\int \dd^2\th \tilde{Q}Q \right)= M' S_{MASS}' + M'' S_{MASS}'' \\
S_{MASS}' &= \frac{1}{2}\left(\tilde{q}f + \tilde{f}q - \tilde{\mu}\mu + \bar{\tilde{q}}\bar{f} + \bar{\tilde{f}}\bar{q} - \bar{\tilde{\mu}}\bar{\mu} \right), \\
S_{MASS}'' &= \frac{1}{2}\left(\tilde{q}f + \tilde{f}q - \tilde{\mu}\mu - \bar{\tilde{q}}\bar{f} - \bar{\tilde{f}}\bar{q} + \bar{\tilde{\mu}}\bar{\mu} \right),
\end{aligned}
\end{equation}
where $M = M' - i M''$ is the four dimensional mass, $M'$ and $M''$ are supposed to be real.


\subsection{Dimensional reduction and topological twist}

To perform the dimensional reduction we take the target space of the four dimensional super Yang--Mills model as a  product $\Sphere^2 \times \Tor_2$ where $\Tor_2$ is a small volume torus. Let the rudimentary torus coordinates be $x^0$ and $x^3$, and $x^1$ and $x^2$ be the coordinates on $\Sphere^2$. Introduce on $\Sphere^2$ a complex structure by identifying $\Sphere^2 = \Compl$. Define 
\begin{equation}
\label{zzbar}
\begin{aligned}
z &= x^1+ix^2 & \bar{z} &= x^1 - i x^2
\end{aligned}
\end{equation}
In the reduced theory the gauge field is the connection of a $G$--bundle $E$ over $\Sphere^2$. Denote this connection by $A = A_\mu \dd x^\mu$, and its curvature by $F_{\mu\nu} = \ep_{\mu\nu}F_{12}$, where $\ep_{\mu\nu}$ is the Levi--Civita tensor in two dimensions, the orientation is such that $\ep_{12} = +1$. The remaining part of the four dimensional gauge field defines two scalars:
\begin{equation}
\begin{aligned}
\l &= \frac{A_0 -  A_3}{2}, & \phi &= \frac{A_0 +  A_3}{2}.
\end{aligned}
\end{equation}

Dimensionally reduced theory has the  usual two dimensional Lorentz symmetry $\SO(2)_L = \UU(1)_L$, as well as the $\R$--symmetry, which is $\UU(1)_\R$. The last one acts on the supermultiplets \Ref{gaugeSUFI} as follows:
\begin{equation}
\begin{aligned}
W_\a(y,\th) &\mapsto \e^{i\vf/2}W_\a (y,\e^{-i\vf/2}\th), & V(x,\th,\bar{\th}) &\mapsto V(x,\e^{-i\vf/2}\th,\e^{i\vf/2}\bar{\th}).
\end{aligned}
\end{equation}
In components it means 
\begin{equation}
\begin{aligned}
A_\mu &\mapsto A_\mu, & \psi &\mapsto \e^{i\vf/2}\psi, & D &\mapsto D.
\end{aligned}
\end{equation}

Usually the $\R$--symmetry is broken by quantum effects. The non--invariance of the fermion measure produces the singlet anomaly. The Atiyah--Singer index theorem claims that this non--invariance is given by the first Chern class $c_1(E) = \frac{F}{2\pi}$, which counts the difference between right handed and left handed zero modes of the two dimensional Dirac operator. When this difference is not zero, the $\R$--symmetry group is broken to a discrete group. However, if we add some matter it is possible to have the unbroken $\R$--symmetry. We will focus in that follows at this situation. 

If the $\R$--symmetry is unbroken, we can perform the topological twist by demanding that each Lorentz rotation on angle $\vf$ should be accompanied by the $\R$--transform with parameter $\vf/2$. In other words we take the diagonal subgroup of the product of the Lorentz group and the $\R$--symmetry group as the new Lorentz group:
\begin{equation}
\UU(1)_L' = \diag\left(\UU(1)_L \times \UU(1)_\R\right)
\end{equation}
Under such a transformation the components of the spinor $\psi$ transform as follows:
\begin{equation}
\begin{aligned}
\psi^+ &\mapsto \e^{i\vf}\psi^+, & \psi^- &\mapsto \psi^-.
\end{aligned}
\end{equation}
Therefore we can write $\psi^+ = \psi_1 + i\psi_2$ where $\psi^\mu$ are components of a vector. Recall that we are in two dimensional Euclidean space and therefore do not distinguish upper and lower indices. Another component, $\psi^-$, is a scalar with respect to infinitesemal rotations. Note however that the complex conjugation is equivalent to the $x^2$--reflection. Therefore if we write $\psi^- = \eta - i \chi$, then $\eta$ will be a scalar, whereas $\chi$ is a pseudoscalar, which changes sign under reflections. 

We define similarly topological twist for the supercharges  and parameters of the supersymmetry transformation:
\begin{equation}
\begin{aligned}
\z^+ &= \frac{v_1 + i v_2}{2}, & \Q_+ &= \Q_1 - i \Q_2 \\
\z^- &= \frac{s - i p}{2}, & \Q_- &= \Q + i \tld{\Q}.
\end{aligned}
\end{equation}
In twisted notations we have simply $\d = s\Q + p \tld{\Q} + v^\mu \Q_\mu$. The commutation rules for twisted supersymmetry operators are
\begin{equation}
\label{SUSYTwistedcommut}
\begin{aligned}
\{ \Q,\tld{\Q}\} &= 0 & \Q^2  &= \G(\phi) & \tld{Q}^2 &= \G(\phi) \\
\{\Q_\mu,\Q_\nu\} &= 2g_{\mu\nu}\G(\l) & \{ \Q,\Q_\mu\} &= -i\cD_\mu & \{\tld{\Q},\Q_\mu \} &= -i\ep_{\mu\nu}\cD^\nu,
\end{aligned}
\end{equation}
where $\G(\a)$ is the gauge transformation with parameter $\a$. 

The reduced and twisted version of algebra \Ref{gaugeSUSY} is given by
\begin{equation}
\label{QtwistGauge}
\begin{aligned}
\Q\phi &= 0 & \tld{\Q}\phi &= 0 & \Q_\mu \phi &= \psi_\mu \\
\Q\l &= \eta & \tld{\Q}\l &= \chi & \Q_\mu \l &= 0 \\
\Q A_\mu &=  \psi_\mu & \tld{\Q} A_\mu &= -\ep_{\mu\nu}\psi^\nu & \Q_\mu A_\nu &= g_{\mu\nu}\eta +\ep_{\mu\nu}\chi\\
\Q\psi_\mu &= -i\cD_\mu \phi & \tld{\Q} \psi_\mu &= - i\ep_{\mu\nu}\cD^\nu \phi & \Q_\mu \psi_\nu  &= - ig_{\mu\nu} [\phi,\l] - \ep_{\mu\nu} (iF_{12} +B)\\
\Q\eta &= i[\phi,\l] & \tld{\Q} \eta &= - B & \Q_\mu \eta &= -i\cD_\mu \l \\
\Q\chi &= B & \tld{\Q}\chi & = i [\phi,\l] & \Q_\mu \chi &= -i\ep_{\mu\nu}\cD^\nu \l\\
\Q B &= i[\phi,\chi] & \tld{\Q}B & = - i[\phi,\eta] & \Q_\mu B &= -i\cD_\mu \chi +i\ep_{\mu\nu}\cD^\nu \eta - i \ep_{\mu\nu}[\l,\psi^\nu].
\end{aligned}
\end{equation}
where we have introduced $B = \frac{-iF_{12}-D}{2}$.

Note that if we get back to original set of fields, that is, plug in back $D$ instead of $B$ we observe that actions of $\Q$ and $\tld{\Q}$ are the same, modulo the change $\eta \leftrightarrow \chi$ and the supplementary Hodge transformation of all vectors: $V_\mu \leftrightarrow {(\star V)}_\mu = \ep_{\mu\nu}V^\nu$. Note also, the if we define $B = \frac{+iF_{12}-D}{2}$ then the action of $\Q_\mu$ gets simpler whereas the actions of $\Q$ and $\tld{\Q}$ get more complicated.

To twist the supersymmetry algebra for the matter fields \Ref{matterSUSY} we arrange the components of $Q$ and $\tilde{Q}$ to $\UU(1)'_L$ spinors:
\begin{equation}
\begin{aligned}
x &= \left(
\begin{array}{c}
\bar{\tilde{q}} \\
-q
\end{array}
\right),  & \xi &= \frac{1}{\sqrt{2}}\left(
\begin{array}{c}
\bar{\tilde{\mu}}^+ \\
\mu^+
\end{array}
\right), & \omega &= \frac{1}{\sqrt{2}} \left(
\begin{array}{c}
\mu^-\\
\bar{\tilde{\mu}}^-
\end{array}
\right), & y &= \frac{1}{2}\left(
\begin{array}{c}
f \\
\bar{\tilde{f}}
\end{array}
\right).
\end{aligned}
\end{equation}
Define also two dimensional Dirac matrices $\g_\mu$ and the chiral matrix $\gG$ as follows:
\begin{equation}
\begin{aligned}
\g_1 &= \left(
\begin{array}{cc}
0 & 1 \\
1 & 0
\end{array}
\right), & \g_2 &= \left(
\begin{array}{cc}
0 & -i \\
i & 0
\end{array}
\right), & \gG &= \g_1\g_2 = \left(
\begin{array}{cc}
i & 0 \\
0 & -i 
\end{array}
\right).
\end{aligned}
\end{equation}
The spinor rotation matrix $\g_{12}$ is defined as usual: $\g_{12} = \frac{1}{4}\left(\g_1\g_2 -\g_2\g_1\right) = \frac{1}{2}\gG$. The twisted superalgebra is given by
\begin{equation}
\label{HypTwist}
\begin{aligned}
\Q x &= \xi & \tld{\Q}x &= \gG \xi & \Q_\mu x &= - \g_\mu \omega \\
\Q \xi &= i \phi x & \tld{\Q} \xi &= -i\gG \phi x & \Q_\mu \xi &= \g_\mu h -i\cD_\mu x  \\
\Q \omega &= h & \tld{\Q}\omega &= - \gG h +i\gG \g^\mu \cD_\mu x & \Q_\mu\omega &= -i\g_\mu \l x \\
\Q h &= i\phi \omega & \tld{\Q}h &= i\gG\phi  \omega -i \gG \g^\mu \cD_\mu \xi & \Q_\mu h &= i \g_\mu \l \xi -i \cD_\mu \omega + i \g_\mu \eta x
\end{aligned}
\end{equation}
\begin{equation}
\label{HypTwistBar}
\begin{aligned}
\Q \bar{x} &= -\bar{\xi} & \tld{\Q}\bar{x} &= \bar{\xi}\gG & \Q_\mu \bar{x} &=\bar{\omega}\g_\mu \\
\Q \bar{\xi} &= i \bar{x}\phi & \tld{\Q} \bar{\xi} &= i \bar{x}\phi\gG & \Q_\mu \bar{\xi} &= \bar{h}\g_\mu  +i \cD_\mu \bar{x}  \\
\Q \bar{\omega} &= \bar{h} & \tld{\Q}\bar{\omega} &= \bar{h}\gG +i \cD_\mu \bar{x} \g^\mu\gG & \Q_\mu\bar{\omega} &= -i\bar{x}\l\g_\mu  \\
\Q \bar{h} &= -i\bar{\omega}\phi & \tld{\Q}\bar{h} &= i \bar{\omega}\phi\gG  +i\cD_\mu \bar{\xi}\g^\mu \gG & \Q_\mu \bar{h} &= -i\bar{\xi} \g_\mu \l -i \cD_\mu \bar{\omega} + i \bar{x}\eta\g_\mu,
\end{aligned}
\end{equation}
where $h = y + \frac{i}{2}\g^\mu \cD_\mu x$ and $\bar{h} = \bar{y} - \frac{i}{2}\cD_\mu \bar{x}\g^\mu$. Note that this substitution simplifies the $\Q$ and $\tld{\Q}$ actions. Had we put $h = y - \frac{i}{2}\g^\mu \cD_\mu x$ and $\bar{h} = \bar{y} + \frac{i}{2}\cD_\mu\bar{x}\g^\mu $ the action of $\Q_\mu$ would become simpler instead. 


\subsection{Twisted superfield and twisted superpotential}

The Fayet--Iliopoulos term \Ref{FI} can appear in the model if the gauge group contains an $\UU(1)$--factor. Let in that follows $G=\UU(N)$. In two dimensions one can introduce such term with the help of the superfield $V(x,\th)$. However, it is not gauge invariant. Another way to do it is to introduce so--called twisted superfield (see \cite{Phases} and references therein). Note that the meaning of the word ``twisted'' in this context has nothing to do with the topological twist. The twisted superfield can be defined as follows. We consider the abelian case. Let $\D_\a$ and $\bar{\D}_{\dot{\a}}$ be covariant superderivatives which commute with supercharges. Define
\begin{equation}
\label{twistSUFI}
\begin{aligned}
\gS = \frac{i}{2}\D_- \bar{\D}_- V &= \phi + \th^+\bar{\psi}^+ + \bar{\th}^+ \psi^+ + \th^+\bar{\th}^+ (-F_{12}-iD) \\
&= \phi + \th^\mu \psi_\mu + \th_1\th_2 \frac{iF_{12} -D}{2} = \phi + \th^\mu \psi_\mu + \th_1\th_2 (iF_{12} + B),
\end{aligned}
\end{equation}
where we have introduced the topological twist for the spinor supercoordinates $\th^+ = \frac{\th_1 + i \th_2}{2}$. All fields are supposed to be functions of $y^\mu$ defined by
\begin{equation}
\label{coord}
\begin{aligned}
y_1 + i y_2 &= z -2i\th^+\bar{\th}^- & y_1 - i y_2 &= \bar{z} - 2i \bar{\th}^+ \th^-.
\end{aligned}
\end{equation}
Note that the twisted superfield is the $\Compl$--analog of the four dimensional $\N=2$ chiral multiplet $\Psi(y,\th)$. Recall that its superspace expansion in four dimensional twisted supercoordinates $\th^I$ is the following:
\begin{equation}
\Psi(y,\th) = \phi(y) + \th^I \psi_I(y) + \th^I\th^J {\left(iF_{IJ}(y) - D_{IJ}(y)\right)}^+ + \dots,
\end{equation}
where $D_{IJ} = D \eta^3_{IJ} + f \eta_{IJ}^+ + f^\dag \eta^-_{IJ}$, $\eta^i_{IJ}$ are  't Hooft symbols, $\eta^\pm = \frac{\eta^1 \pm i \eta^2}{2}$, $D$ and $f$ are the auxiliary fields for $\N=1$ gauge and chiral multiplets in four dimensions, and ${(\dots)}^+$ means the self--dual part. It would be interesting to see if this four dimensional superfield can be obtained as a sort of ``twisted'' $\N=1$, $d=6$ vector multiplet.

Introduce the complex parameter
\begin{equation}
\label{tau}
\tau = ir + \frac{\gTh}{2\pi}.
\end{equation}
Its four dimensional analog is $\tau_4 = \frac{\gTh}{2\pi} + \frac{4\pi i}{g^2}$. The analog of $r$ is, therefore, $\frac{4\pi}{g^2}$. It follows that weak interacting regime in the $\Quat$--case corresponds to $r \to \infty$ regime in the $\Compl$--case. Define also the twisted Grassman measure $\dd^2 \vec{\th} =  \dd \th_2 \dd \th_1$. Then the topological action \Ref{Stop} together with the Fayet--Iliopoulos term \Ref{FI} is generated by two dimensional F--term:
\begin{equation}
\label{Fmicro}
2\Im \left(\tau \int \dd^2 \vec{\th} \gS \right) = S_{top} + S_{FI}.  
\end{equation}
Let us make a remark. In Euclidean space the complex conjugation raises and lowers indices. Thus it corresponds to the exchange $\psi^+ \leftrightarrow \psi^-$. It follows that the complex conjugated twisted multiplet looks like
\begin{equation}
\label{twistSUFIbar}
\begin{aligned}
\bar{\gS} = \frac{i}{2}\D_+\bar{\D}_+ V &= \l + \th^-\bar{\psi}^- + \bar{\th}^-\psi^- + \th^-\bar{\th}^- (F_{12}-iD) \\
&= \l + \th \eta + \tilde{\th}\chi + \th\tilde{\th}\frac{iF_{12} + D}{2},
\end{aligned}
\end{equation}
where $\th^- = \frac{\th - i\tilde{\th}}{2}$, and all component fields are functions of $y_1 + i y_2 = z + 2i \th^+\bar{\th}^-$ and $y_1 - iy_2 = \bar{z} + 2i\bar{\th}^+\th^-$. Note also that the abelian version of the action \Ref{Sgauge} reduced to two dimensions can be written in two dimensions as D--term:
\begin{equation}
\label{SgaugeSigmaSigmaBar}
S_{gauge} = \int \dd^2 \th \dd^2 \bar{\th} \bar{\gS}\gS.
\end{equation}

In quantum theory the F--term gains corrections, perturbative and non--perturbative. If the supersymmetry remains unbroken in the quantum level, the most general form of the F--term is given by generalization of \Ref{Fmicro}:
\begin{equation}
\label{twistedSup}
2\Im\left(\int\dd^2 y \dd^2\vec{\th} \frac{1}{2\pi i}W_0(\gS,\tau) \right),
\end{equation}
where $W_0(\gS,\tau)$ is the twisted superpotential. In the microscopic theory it is linear function of $\gS$:
\begin{equation}
\label{Wclass}
W^{class}_0(\gS,\tau) = \pi i \tau \Tr \gS.
\end{equation}

The perturbative corrections to the twisted superpotential can be interpreted as the renormalization group flow for the complex parameter \Ref{tau}. Typically it has the following form (we have $\frac{1}{\pi i}$ factor instead of traditional $\frac{1}{2\pi i}$ since we have chosen non--usual normalization in the action \Ref{Sgauge}):
\begin{equation}
\label{RGflow}
\tau(\gL_1) = \tau(\gL_2) + \frac{\beta}{\pi i} \ln \frac{\gL_1}{\gL_2},
\end{equation}
where $\gL$ is the dynamically generated scale. $\beta$ is first term of the $\beta$--function expansion. For $\UU(N)$ theory with $N_F$ matter multiplets in the fundamental representation  and $N_{\tilde{F}}$ in the antifundamental we have $\b = N_{\tilde{F}} - N_F$ \cite{Phases}. The $\R$--symmetry is broken down to a discrete group: $\UU(1)_\R \mapsto \Integer/ \beta \Integer$. The only case when it survives is $\beta =0$, which is possible if $N_F = N_{\tilde{F}}$. The theory we are interested in is, therefore, conformal at the quantum level.

The rest of the paper is devoted to the explicit computations of the quantum corrections to the twisted superpotential for this case.


\section{CohFT features of the model}
\label{Coh}


\subsection{Action is $\Q$--exact}

Topologically twisted theory possess two scalar  fermionic operators: $\Q$ and $\tld{\Q}$. Strictly speaking the last operator is pseudoscalar, but this difference is not essential. The action is supersymmetric, that is, in particular, $\Q$ and $\tld{\Q}$ closed. Therefore it has good chances to be $\Q$ and $\tld{\Q}$ exact. Indeed, the computation shows that:
\begin{equation}
\begin{aligned}
S_{gauge} &= \Q \Psi_{gauge}, & S_{FI} &= \Q \Psi_{FI}, & S_{matter} &= \Q \Psi_{matter} \\
\Psi_{gauge} &= \tld{\Q}V_{gauge}, & \Psi_{FI} &= \tld{\Q}V_{FI}, & \Psi_{matter} &= \tld{\Q} V_{matter},
\end{aligned}
\end{equation}
where
\begin{equation}
\label{PsiV}
\begin{aligned}
\Psi_{gauge} &= -i\chi(F_{12}-iB) - i\l\cD_\mu \psi^\mu  - i \eta [\phi,\l] & V_{gauge} &= -i\l F_{12} + \eta \chi \\
\Psi_{FI} &= 2r \chi & V_{FI} &= 2r \l \\
\Psi_{matter} &= i(\bar{\xi}\l x + \bar{x}\l \xi) - 2i \bar{x}\g_{12}\chi x  & V_{matter} &= -2i \bar{x}\l \g_{12}x - \bar{\omega}\gG \omega \\
&\phantom{=} + (\bar{h}+i\cD_\mu \bar{x}\g^\mu )\omega + \bar{\omega}(h-i\g^\mu \cD_\mu x)
\end{aligned}
\end{equation}
Therefore the action can be written as follows: 
\begin{equation}
S = S_{gauge} + S_{FI} + S_{matter} = \Q\tld{\Q}\left( V_{gauge} + V_{FI} + V_{matter} \right),
\end{equation}
which shows that we deal with $\N_T=2$ cohomological theory (recall that both supercharges anticommute). This is not true for the four dimensional mass term \Ref{4dMass}. Instead we have:
\begin{equation}
\begin{aligned}
S'_{MASS} &= \Q \Psi_{MASS}' = \tld{\Q}\Psi_{MASS}'' \\
S''_{MASS} &= - \Q \Psi_{MASS}'' = \tld{\Q}\Psi_{MASS}' \\
\Q_\mu \left(\bar{x}\g_\nu x\right) &= i g_{\mu\nu}\Psi''_{MASS} + i \ep_{\mu\nu}\Psi'_{MASS}.
\end{aligned}
\end{equation}
Therefore this term is $\Q$ and $\Q_\mu$ or $\tld{\Q}$ and $\Q_\mu$ exact, but never both. Since $\bar{x}\g_\mu x$ is gauge invariant we conclude using \Ref{SUSYTwistedcommut} that $\Q_\mu \Psi'_{MASS} = \Q_\mu \Psi''_{MASS} = 0$.

Moreover it is straightforward to check that 
\begin{equation}
V_{matter} = \frac{1}{4}\Q_\mu \left(\bar{x}\gG \g^\mu \omega + \bar{\omega}\gG \g^\mu x\right).
\end{equation}
It follows that $\Q_\mu V_{matter} = 0$ and therefore
\begin{equation}
\Q_\mu \Psi_{matter} = \{ \Q_\mu,\tld{\Q} \} V_{matter} = -i\ep_{\mu\nu}\pd^\nu V_{matter}.
\end{equation}
Since $\Q_\mu V_{FI} = 0$, the same is true for the Fayet--Iliopoulos term:
\begin{equation}
\label{QmuPsiFI}
\Q_\mu \Psi_{FI} = -i\ep_{\mu\nu}\pd^\nu V_{FI}.
\end{equation}
Finally one can check that 
\begin{equation}
\label{PsiGauge}
\Psi_{gauge}' = \Psi_{gauge} + i \pd_\mu (\l \psi^\mu)  = - \frac{1}{2}\Q_\mu \left(\chi \ep^{\mu\nu}\psi_\nu + \eta\psi^\mu \right)
\end{equation}
It follows that 
\begin{equation}
\label{QmuPsiGauge}
\begin{aligned}
\Q_\mu \Psi_{gauge}' &= 0 & \Q_\mu \Psi_{gauge} &= -i\pd_\rho \Q_\mu (\l\psi^\rho) = i\ep_{\mu\rho}\pd^\rho\left(\l(iF_{12}+B)\right).
\end{aligned}
\end{equation}


\subsection{Gauge fixing}

As probably all theories, whose action is $\Q$--exact for a fermionic scalar operator $\Q$, the model in question can be obtained by the gauge fixing for an appropriate action. Recall how it works \cite{BaulieuD=4N=1}.

Consider the following ``topological'' action:
\begin{equation}
\label{Stop}
S_{top} = \gTh c_1(E) = \frac{\gTh}{2\pi} F_{12},
\end{equation}
where $\gTh$ is the two dimensional instanton angle. For this action to be non zero it is necessary to have at least one generator with non vanishing trace in the gauge group Lie algebra. In other words, the gauge group has to contain at least one $\UU(1)$ factor. In such a  situation the Fayet--Iliopoulos term \Ref{FI} is always acceptable. To be specific, in that follows we consider the model for $G = \UU(N)$.

The topological action equals $\gTh k$, where $k$ is the winding number for the gauge field configuration. Therefore it is invariant not only with respect to  usual gauge transformations of the connection $A_\mu \mapsto A_\mu - \cD_\mu \a$, but also under small generic deformations: $A_\mu \mapsto A_\mu + \a_\mu$, provided $A_\mu$ and $A_\mu + \a_\mu$ belong to the same homotopic class.

To fix both gauge invariances we have to introduce a BRST (BV) operator $\Q$ as well as a set of ghosts, antighosts and gauge fixing conditions. Denote the small deformation ghost by $\psi_\mu$, the Lagrange multiplier by $B$ and the antighost by $\chi$. Then the BRST operator acts as follows:
\begin{equation}
\begin{aligned}
\Q A_\mu &= \psi_\mu, &
\Q\chi &= B.
\end{aligned}
\end{equation}
Let the gauge fermion be $\Psi = -i\chi F_{12}$. It produces term $-iB F_{12}$ in the action, and therefore the gauge fixing condition is $F_{12}=0$, the flat connection. Consider the kinetic term for ghosts:
\begin{equation}
\frac{1}{2}\ep^{\mu\nu} \left(\cD_\mu\psi_\nu - \cD_\nu \psi_\mu \right).
\end{equation}
It has following symmetry: $\psi_\mu \mapsto \psi_\mu - \cD_\mu \z$, where $\z$ is a fermionic gauge parameter.Indeed, the variation of the kinetic term is given by $F_{12}\z = 0$ thanks to the gauge fixing condition for the connection. Therefore there is another gauge symmetry to be fixed. Denote corresponding ghost by $\phi$, the antighost by $\l$ and the Lagrange multiplier by $\eta$. The extended action of the BRST operator is given precisely by the first column of \Ref{QtwistGauge}. Note also that if we choose the gauge fixing condition for $\psi_\mu$ as $\cD_\mu \psi^\mu = 0$, then the gauge fermion will be $\Psi = -i\chi F_{12} + \l \cD_\mu \psi^\mu$, which is up to a potential $-\chi B - i\eta[\phi,\l]$ match with \Ref{PsiGauge}. This potential does not affect on the singularities structure of the action \cite{TQFT}, and we conclude that this gauge fixed action is equivalent to the action of our model. 

The BRST operator introduced in this way is not nilpotent. Instead, as we see in \Ref{SUSYTwistedcommut}, it satisfies $\Q^2 = \G(\phi)$. To get really nilpotent operator we have to fix the rest of the gauge freedom. To this extent we introduce the ghost $c$, antighost $\bar{c}$ and the Lagrange multiplier $b$. The full BRST operator acts as follows:
\begin{equation}
\begin{aligned}
\Q A_\mu &= \psi_\mu  - i \cD_\mu c & \Q \psi_\mu &= -i\cD_\mu \phi + i\{ c, \psi_\mu\} \\
\Q \phi &= i \{c,\phi \}  & \Q c &= \frac{i}{2}\{ c,c\} - \phi \\
\Q \bar{c} &= b & \Q b &= 0 \\
\Q \chi &= B + i \{c,\eta \} & \Q B &= i[\phi,\chi] +i[c,B] \\
\Q \l &= \eta + i[c,\l] & \Q \eta &= i[\phi,\l] + i \{c,\eta \}. 
\end{aligned}
\end{equation}
It is straightforward to check that it is nilpotent. 

Let us also describe the mass for the matter multiplet. As we have seen, the four dimensional mass term \Ref{4dMass} is $\Q$--exact, and hence appears as the deformation of the gauge fermion. Two theories (with and without this term) are equivalent. Another way to introduce the mass consists of a deformation of the BRST operator. The action remains BRST--exact, but the BRST operator itself is deformed. The simplest way to produce the mass is to perform the formal shift: 
\begin{equation}
\label{shift}
\begin{aligned}
\phi &\mapsto \phi + m & \l &\mapsto \l + m,
\end{aligned}
\end{equation}
where $m$ is the mass. The action gains the following contribution:
\begin{equation}
\label{Smass}
S_{mass} = -2m^2 \bar{x}x - 2m \bar{x}\phi x - 2m \bar{x}\l x - 2im \bar{\xi}\xi - 2i m\bar{\omega}\omega.
\end{equation}
The BRST algebra for the gauge multiplet remain unchanged, but gets deformed for the matter fields:
\begin{equation}
\begin{aligned}
\Q_m x &= \xi & \Q_m &= - \bar{\xi}\\
\Q_m \xi &= i \phi x + i m x & \Q_m \bar{\xi }&= i\bar{x} \phi + i m \bar{x}\\
\Q_m \omega &= h & \Q_m \bar{\omega} &= \bar{h}\\
\Q_m h &= i\phi \omega + im\omega & \Q_m \bar{h} &= -i\bar{\omega} \phi - im \bar{\omega}.
\end{aligned}
\end{equation}
We see that $\Q_m^2 = \G(\phi) + \F(m)$, where $\F(m)$ is the flavor group action: $\F(m)Q = im Q$ and $\F(m)\tld{Q} = -im \tld{Q}$. Note that  $\F(m)V = 0$. Hence the deformed BRST operator matches with undeformed one when it acts on the gauge multiplet: $\Q V = \Q_m V$. To prove that the action remains $\Q_m$--exact we notice that 
\begin{equation}
\begin{aligned}
\Q_m \Psi_{matter} &= \Q \Psi_{matter} - 2im \bar{x}\l x -2im \bar{\omega}\omega \\
\Q_m \left(\bar{\xi} x + \bar{x}\xi \right) &= 2i\bar{x}\phi x + 2im\bar{x}x-2\bar{\xi}\xi.
\end{aligned}
\end{equation}
It follows that
\begin{equation}
S_{matter} + S_{mass} = \Q_m\left(\Psi_{matter} +\Psi_{mass}\right),
\end{equation}
where $\Psi_{mass} = im(\bar{\xi} x + \bar{x}\xi)$. Note that it can be obtained from \Ref{PsiV} by the formal shift \Ref{shift}.

Tilded and untilded chiral multiplets transform in \Ref{HypTwist} and \Ref{HypTwistBar} independently. We have put them into for of Dirac spinors to shorten formulae. Also they enter separately into the Lagrangian. The only term that mixes them is the four dimensional mass \Ref{4dMass}. If we delete it, we can consider two multiplets independently. In particular, they can have different two dimensional masses: $\F Q = i m Q$ and $\F \tilde{Q} = -i\tilde{m}\tilde{Q}$.


\section{Ground states of the model}
\label{vortex}

In this section we describe classical vacua of the model in question as well as certain deformation of the model. The deformation is needed for the following reason. We are interested in the non--perturbative corrections to the twisted superpotential. We would like to think of them as of ``small'' corrections, even smaller that perturbative ones. It corresponds to the regime $|r|\gg 0$. But in this regime the initial theory does not possess a Coulomb branch. However we can deform it, introducing an asymmetry between left and right movers in twisted theory, or, equivalently, between $Q$ and $\tilde{Q}$ in the untwisted one. The deformed model do have a Coulomb branch which is consistent with $r \neq 0$. The price we pay is that we lose three quarters of the supersymmetry.


\subsection{Vacua}

Consider the full action of the model with $N_F$ flavors of the untilded matter and $N_{\tilde{F}}$ flavors of the tilded one. Recall that their numbers and their masses do not necessarily match. In principle, even the representations of $Q$ and $\tilde{Q}$ may be independent, and not be dual to each other. Also we can consider more than two different representations. The action can be written as follows:
\begin{equation}
\label{Sfull}
S_{full} = S_{top} + S_{gauge} + S_{FI} + S_{matter} + S_{mass} = \tau F_{12} + \Q_m \Psi_{full}
\end{equation}
Using the $\Q_m$--exactness of this action we take more general expression for the gauge fermion than \Ref{PsiV}. Namely, let
\begin{equation}
\label{PsiFull}
\begin{aligned}
\Psi_{full}&= - \chi\left(iF_{12} -2r + 2i x \g_{12} \bar{x} + t_g B\right) + i \psi^\mu \cD_\mu \l -iA_1\eta[\phi,\l] -iA_2 m (\bar{\xi} x + \bar{x}\xi) \\
&\phantom{=} + i A_3 (\bar{\xi}\l x +\bar{x}\l \xi)  + (t_m \bar{h}+i\cD_\mu \bar{x}\g^\mu)\omega + \bar{\omega}(t_m h- i \g^\mu \cD_\mu x),
\end{aligned}
\end{equation}
where $A_1$, $A_2$, $A_3$, $t_g$ and $t_m$ are arbitrary constants. The vacuum expectation of any $\Q_m$--exact quantity is independent of them. The initial model corresponds to $A_1=A_2=A_3=t_g=t_m=1$. For the sake of brevity we omit the summation on flavor indices and the indices themselves. For example $2x\g_{12}\bar{x}$ should be read as $ i \sum_{f=1}^{N_F} q_{f,l} \bar{q}_{f,m}  -i \sum_{f=1}^{N_{\tilde{F}}} \tilde{q}_{f,m} \bar{\tilde{q}}_{f,l}$, where $l,m$ are color indices. 

Let us integrate out auxiliary fields $h$, $\bar{h}$ and $B$. The bosonic part of the action is given by the following expression:
\begin{equation}
\label{Sboson}
\begin{aligned}
S_{boson} &= \frac{1}{4t_g} {\left(iF_{12}-2r +2ix\g_{12}\bar{x} \right)}^2 + \cD_\mu \phi \cD^\mu \l  - \frac{1}{2t_m} \left( \cD_\mu \bar{x}\g^\mu\right)\left(\g^\nu \cD_\nu x\right) \\
&\phantom{=} + A_1{[\phi,\l]}^2 - 2A_2m^2 \bar{x}x  -2A_2m\bar{x}\phi x -2A_3m\bar{x}\l x - A_3\bar{x}(\phi \l + \l \phi )x.
\end{aligned}
\end{equation}
Now we can describe the vacua of the theory. When $r=0$ the vacuum is given by the following equations: $x = 0$, $\bar{x}=0$, $[\phi,\l] = 0$. The $A_1$--term implies that in this situation $\phi$ and $\l$ are diagonal (as it follows from \Ref{twistSUFI} and \Ref{twistSUFIbar} we have to identify $\bar{\phi} = \l$, as it would be had we started from Euclidean four dimensional space, or had we performed the Wick rotation in the torus on which we compactify the theory). The gauge group is broken down to its maximal torus: $\UU(N)\mapsto {\UU(1)}^N$ and the theory is in the Coulomb branch. 

However, as we have mentioned in the beginning of this section, when $r = 0$ the non--perturbative corrections are not really small. Moreover as we shall see later, this condition is not compatible with the localization technique. See also \cite{NikitaLosevTauBar} for further explanation. 

Let us, therefore, study cases when $r\neq 0$. In components the first term in the boson part of the action \Ref{Sboson} is given by the following expression ${(iF_{12} -2r  - \bar{q}q + \bar{\tilde{q}}\tilde{q})}^2$. We see, that if $r > 0$, the vacuum energy is zero if (let us stress that this is only sufficient condition) $N_{\tilde{F}}=N$, $\tilde{q}_{f,l} = \sqrt{2r} \d_{f,l}$ and $q_{f,l} = 0$. If $r < 0$ then $Q$ and $\tilde{Q}$ are interchanged: $N_F = N$, $q_{f,l} = \sqrt{2|r|}\d_{f,l}$ and $\tilde{q}_{f,l} = 0$. Let us in that follows chose $r > 0$. We see, that some of matter multiplets acquire a non--zero vacuum expectation. It follows that if we wish to have zero vacuum energy, we must put $\phi_{lm} = \d_{l,m} a_l$, where $a_l = - \tilde{m}_l$. In particular, if the mass of $\tilde{Q}$ is zero, it implies $\phi = 0$. The group of the global symmetry of the theory is $\UU(N)_G\times \UU(N_{\tilde{F}})_{\tilde{F}} \times \UU(N_F)_F$, where first factor is the global gauge transformation (gauge transformations at infinity), whereas the rest is the flavor group for $\tilde{Q}$ and $Q$ respectively. If $\tilde{Q}$ acquires non--zero vevs, this group is broken down to $\UU(N)'\times \UU(N_F)_F$, where first factor is diagonal part of the product of gauge group and $\tilde{Q}$ flavor group. The theory is in the color--flavor locking phase, where it has $N$ separated vacua which are permuted by the Weyl group of the gauge group, and there is no Coulomb branch. We can also put $N_{\tilde{F}} > N$. In this situation the Higgs branch appears, but we still can not find the Coulomb branch is such a way.

To obtain the Coulomb branch we must eliminate $A_2$ and $A_3$  terms for $\tilde{Q}$. Then non--zero vevs of $\tilde{Q}$ will be compatible with non--zero vevs of $\phi$ and $\l$. In such a way we lose three quarters of the supersymmetry. Namely, the action is not invariant any more with respect to neither $\tld{\Q}$ nor $\Q_\mu$. Only $\Q = \frac{\Q_- + \bar{\Q}_-}{2}$ survives. On the other hand we have no more such a severe restriction imposed on $\phi$. The only condition is given by $A_1$--term, and we recover the Coulomb branch. Note that the only terms which break the supersymmetry are those which contain the components of $\tilde{Q}$. 

The topological sector of the deformed model is the same as the topological sector of the initial model. Since our main assumption is that the F--term is fully defined by the topological sector, it is natural to expect that twisted superpotential computed for the deformed model gives the answer for the undeformed one. It would be interesting to check this statement by the direct computations. However, this is beyond the scope of the present paper. 
  

\subsection{Vortices}

Now let us move $t_g$ and $t_m$. Consider the limit $t_g\to 0$ and $t_m\to 0$. \Ref{Sboson} shows that the functional integral for the vacuum expectation of an observable localizes on solutions for following equations:
\begin{equation}
\label{Bogomol}
\begin{aligned}
iF_{12} + \tilde{q} \bar{\tilde{q}}  - A_4 q\bar{q}&= 2r, &
\cD_{\bar{z}} \bar{\tilde{q}} &= 0, &
\cD_z q  &= 0
\end{aligned}
\end{equation}
modulo the gauge transformation. Here we used once again the topological character of the theory and introduced an arbitrary constant $A_4$. We can now consider the limit $A_4\to 0$. In this limit the first two equations become the two dimensional Bogomol'ny equations. Their four dimensional analog in this context is the Seiberg--Witten monopole equations.

The solutions for the two--dimensional Bogomol'ny equations are known as vortices. Denote the moduli space of vortices as $\V$.  Vortices are classified by the vortex number, which is the first Chern class $c_1(E)$. The dimensions of the vortex  moduli space with fixed value $k$ of $c_1(E)$ is equal to $2Nk$. Denote it by $\V_k$. We have
\begin{equation}
\label{vortexModuli}
\begin{aligned}
\V &= \bigoplus_{k=1}^\infty \V_k,  & \dim \V_k &= 2Nk, & \V_k &= \Compl \times \tld{\V}_k,
\end{aligned}
\end{equation}
where $\Compl$ in the last equality describes the center mass position of $k$ vortices and $\tld{\V}_k$ describes its internal structure. For $k=0$ the moduli space consists of a single point $A_\mu = 0$, $\tilde{q}_{f,l} = \sqrt{2r} \d_{f,l}$.

The moduli space $\V_k$ can be described with the help of a finite dimensional model \cite{VortBranes,NonAbVort}. The construction is the following. Consider a complex $k\times k$ matrix $C$ and another complex $N \times k$ matrix $I$. One can show that the moduli space is given by the K\"ahler quotient 
\begin{equation}
\label{Vk}
\begin{aligned}
\V_k &= \mu^{-1}(2r) / \UU(k), & \mu &= [C^\dag, C] + I I^\dag.
\end{aligned}
\end{equation}
The action of $\UU(k)$ is Hamiltonian, the corresponding moment map is $\mu$. We have simply $C \mapsto g C g^{-1}$ and $I \mapsto gI$, $g \in \UU(k)$. Note that the moment can be obtained form the first line of the Bogomol'ny equations \Ref{Bogomol} by formal replacement $A_{\bar{z}} \mapsto C$ and $\tilde{q} \mapsto I$. 

The last equation in \Ref{Bogomol} describes the solutions for the two dimensional Weyl equation in the vortex background. As it follows from the Atiyah--Singer index theorem, there are exactly $k$ solutions. To select one of them we need to introduce a vector $w$ belonging to $k$--dimensional complex vector space. As in the four dimensional theory (see \cite{MyThesis}) the statistics of this parameter should be fermionic. Therefore  $w \in \Pi \Compl^k$.


\section{Model in $\gO$--background}
\label{OmegaBack}


\subsection{Definition}

We will be interested in the partition function of the model in the Coulomb phase. The vacuum expectations for $\phi$ belong to the Cartan subalgebra of the gauge group: $\phi_{lm} = \d_{l,m}a_l$. The partition function can be written as follows: $Z(a) = \< 1 \>_a$. However, this quantity considered ``as is'' is not useful, since it diverges. Indeed, the theory is Poincar\'e invariant in two dimensions. Since ``1'' is also translation invariant, the full expression is proportional to the volume of two dimensional space.

To regularize this divergence, we have to spoil the translation invariance. It can be done by introducing so--called $\gO$--background in the four dimensional space, and then compactify theory to two dimensions in this background \cite{SmallInst,SWandRP}. 

The anzatz for the $\gO$--background is the following:
\begin{equation}
\label{metric}
\dd s_4^2 = G_{IJ}\dd x^I \dd x^J = {\left(\dd x^0\right)}^2 - {\left(\dd x^3 \right)}^2 - g_{\mu\nu} \left(\dd x^\mu + V^\mu_a \dd x^a \right)\left(\dd x^\nu + V^\nu_a \dd x^a \right),
\end{equation}
where $a = 0,3$ and $V^\mu_a = \gO_a^{\mu\nu} x_\nu$, where $\gO_a^{\mu\nu}$ is a two--dimensional Lorentz rotation matrix. Denote 
\begin{equation}
\label{OmegaOmegaBar}
\begin{aligned}
V^\mu &= \frac{V_0^\mu+V_3^\mu}{2} = \gO^{\mu\nu} x_\nu & \bar{V}^\mu &= \frac{V^\mu_0 - V^\mu_3}{2} = \bar{\gO}^{\mu\nu} x_\nu \\
\gO^{\mu\nu} &=  \frac{\gO^{\mu\nu}_0 + \gO^{\mu\nu}_3}{2}& \bar{\gO}^{\mu\nu}  &= \frac{\gO^{\mu\nu}_0 - \gO^{\mu\nu}_3}{2}.
\end{aligned}
\end{equation}
The determinant of this metric is $\det_{IJ} G_{IJ} = -1$. The only non--zero Ricci coefficients are $\g_{a,\mu\nu} = - \g_{a,\nu\mu} = - \gO_{a,\mu\nu}$. It follows that the metric is flat when $\gO$ and $\bar{\gO}$ commute. Since we are in two dimensions, we can introduce two parameters $\eps$ and $\bar{\eps}$ defined as follows: $\gO_{\mu\nu} = \eps\ep_{\mu\nu}$ and $\bar{\gO}_{\mu\nu} = \bar{\eps}\ep_{\mu\nu}$. 

To see what changes in the $\gO$--background, consider the following derivative $\cD_+ = \frac{1}{2}\left(\cD_0 + \cD_3 \right)$. When the metric is constant, after compactification we have $\cD_+ = \phi$. In the $\gO$--background one has to replace this derivative by the covariant one, which is given by $\frac{1}{2}\left(e_0^I \cD_I + e_3^I \cD_I \right)$, where $e_a^I$ is the vierbein for the metric \Ref{metric}. We have: $e_a^b = \d_a^b$ and $e_a^\mu = - V^\mu_a$. It follows that $\phi$ (and, by same arguments, $\l$) become differential operators:
\begin{equation}
\label{shiftPhiL}
\begin{aligned}
\hat{\phi} &= \frac{1}{2}\left(e_0^I \cD_I + e_3^I \cD_I \right) = \phi - V^\mu \cD_\mu & &\mbox{and} & \hat{\l} &= \frac{1}{2}\left(e_0^I \cD_I - e_3^I \cD_I \right) = \l - \bar{V}^\mu \cD_\mu.
\end{aligned} 
\end{equation}
Looking at the equation \Ref{SUSYTwistedcommut} we conclude that the BRST operator in the $\gO$--background gets deformed in such a way that it satisfies  $\Q'^2 = i \hat{\phi} = i\phi - i\gO_{\mu\nu} x^\nu \cD^\mu$. Once again using formulae \Ref{SUSYTwistedcommut} we see that a good candidate for the deformed operator is \cite{SWfromInst}
\begin{equation}
\label{Qprime}
\Q' = \Q + \gO^{\mu\nu}x_\nu \Q_\mu.
\end{equation}


\subsection{Deformed action}

Now let us focus on the gauge multiplet. Using equations \Ref{QmuPsiFI} and \Ref{QmuPsiGauge} we conclude that $\Q'\Psi'_{gauge} = \Q\Psi_{gauge}' = S_{gauge}$. Thus all changes are caused by the Fayet--Iliopoulos term. To figure them out we first notice that 
\begin{equation}
\label{Q'PsiFI}
\Q' \Psi_{FI} = 2r B -2ir \eps x^\mu \cD_\mu \l = \Q \Psi_{FI} + 4ir \eps \l.
\end{equation}
However the additional term $4ir\eps \l$, and therefore, the whole deformed action is not real. It indicates that some supplementary terms appear. These terms should contain $4i\bar{\eps}r\phi$. A reasonable try is $\Psi_{FI}' = 2r \chi + 2r\bar{\eps}x^\mu\psi_\mu$. Then 
\begin{equation}
\label{Q'PsiFI'}
\Q'\Psi_{FI}' = 2r B + 4ir\eps\l + 4ir \bar{\eps}\phi -2r\eps\bar{\eps} x^2 (iF_{12}+B).
\end{equation}
Now the action is real. Note that the modification of $\Psi_{FI}$ can be interpreted as shift defined by equation \Ref{shiftPhiL} put into formula \Ref{PsiV}.

An important observation is  that additional terms can be interpreted as the following superspace dependence of the complex parameter \Ref{tau}:
\begin{equation}
\label{tau(x,th)}
\tau(x,\th) = \frac{\gTh}{2\pi} + ir \left( 1 + 2\bar{\eps}\left(\ep_{\mu\nu}\th^\mu\th^\nu  +i \eps\bar x^2\right)\right).
\end{equation}
All modifications are proportional to $r$, and therefore are absent when $r=0$.

In the $\gO$--background the supersymmetry is broken. $\Q'$ is the only survived supercharge. In coordinates \Ref{coord} it takes the following form (here we have restricted it to the subspace $\th = \tilde{\th} = 0$):
\begin{equation}
\label{Cartan}
\Q' = \th^\mu \frac{\pd}{\pd y^\mu} - i \gO^{\mu\nu}y_\nu \frac{\pd}{\pd \th^\mu}.
\end{equation}
If we formally identify $\th^\mu = \dd x^\mu$, then the differential operator \Ref{Cartan} becomes the Cartan differential $\Q' = \dd + i_V$, where $V^\mu = - i\eps \ep^{\mu\nu}x_\nu$ is the vector field for the Lorentz rotation. The basic property of this operator is that it annihilates the superspace dependent complex parameter: 
\begin{equation}
\label{Q'tau}
\Q' \tau(y,\th) = 0.
\end{equation}

Now we can establish a proposal for the twisted superpotential. To this extent  we compute the vacuum expectation of ``1'' in the Coulomb phase. We can compute it in two steps: first we integrate out all high--energy modes, and then we integrate the rest. In the Coulomb phase the only massless modes are those which belong to the Cartan subalgebra, that is, the diagonal elements of \Ref{gaugeSUFI}. They can be packaged to \Ref{twistSUFI} and \Ref{twistSUFIbar}. Thus we can write (``$c.c.$'' stands for complex conjugated)
\begin{equation}
\label{Partition}
\begin{aligned}
\< 1\>_a &= \int \D \gS \D \bar{\gS} \e^{i\int \dd^2 y \dd^2\th \frac{1}{2\pi i}W(\gS(y,\th),\tau(y,\th)) + c.c} \\
&= \e^{i\int \dd^2 y \dd^2 \th \frac{1}{2\pi i}W(a,\tau(y,\th)) + c.c.} = \e^{\frac{i}{\eps}W(a,\tau,-i\eps) + c.c.} = {\left|\e^{\frac{i}{\eps}W(a,\tau,-i\eps)} \right|}^2.
\end{aligned}
\end{equation}
The function $W(a,\tau)$ in the righthand side consists of three parts: classical, which is given by equation \Ref{Wclass}, the perturbative part $W_{pert}$, which is entirely defined by the 1--loop expression, and the non--perturbative part $W_{vort}$, due to vortices. In that follows we rotate the parameter of the $\gO$--background in the complex plane: $\eps\mapsto i\eps$.


\section{Computation of twisted superpotential}
\label{Super}


\subsection{Perturbative part}

Let us first compute the perturbative contribution to the twisted superpotential. When $\gO$--background is absent,  the perturbative contribution to the partition function is trivial, since the theory is topological. Fermionic determinant compensates the bosonic one. In the presence of $\gO$--background things change \cite{SmallInst}.

The off--diagonal part of the vector multiplet gains mass thanks to the non--zero vacuum expectation of $\phi$. It happens thanks to terms of form ${[\phi,V]}_{lm} = (a_l - a_m)V_{lm}$. In the $\gO$--background $\phi$ becomes differential operator. In the complex coordinates \Ref{zzbar} shift \Ref{shiftPhiL} can be rewritten as follows: $\hat{\phi} = \phi - \eps(z\pd_z - \bar{z}\pd_{\bar{z}})$. Hence the Higgs mass becomes a differential operator as well:
\begin{equation}
\label{V_lm_Mass}
{[\hat{\phi},V]}_{lm} = \left(a_l - a_m - \eps\left(z\pd_z - \bar{z}\pd_{\bar{z}}\right)\right)V_{lm}.
\end{equation}

To find the perturbative contribution to the partition function we inspect the Yukawa interactions as well as gauge coupling to the Higgs field in the full action \Ref{Sfull}. Let us represent the fields of the gauge multiplet as follows $V_{lm} = \sum_{ij}^\infty v_{lm,ij}z^i \bar{z}^j\e^{-|z|}$. Here the summation on $i,j$ takes into account the Lorentz properties of the component fields of vector multiplet. For example for scalar fields the summation is understood as $\sum_{i,j}^\infty = \sum_{i=0}^\infty \sum_{j=0}^\infty$ whereas for $A_z$ it is $\sum_{i,j}^\infty = \sum_{i=1}^\infty\sum_{j=0}^\infty$, and so on. The relevant part of the boson--fermion determinant ratio is
\begin{equation}
\label{Zpert^gauge}
Z_{pert}^{gauge} = {\left|\e^{\frac{1}{\eps}W_{pert}^{gauge}(a,\eps)} \right|}^2= {\left|\prod_{l\neq m}^N\prod_{i,j=0}^\infty\frac{a_l -a_m - \eps(i-j)}{a_l-a_m -\eps(i-j+1)}\right|}^2 = {\left|\prod_{l\neq m}^N \prod_{i=0}^\infty (a_l-a_m + i \eps ) \right|}^2.
\end{equation}
Same reasoning for the matter multiplets leads to the following contribution of the $Q$ and $\tilde{Q}$ (only $\omega$--terms are relevant). 
\begin{equation}
\label{Zpert^matter}
Z_{pert}^{matter} = {\left|\e^{\frac{1}{\eps}W_{pert}^{matter}(a,\eps)} \right|}^2= {\left|\prod_{l=1}^N \prod_{i=0}^\infty \left(a_l + m - \eps\left(i + \frac{1}{2}\right)\right)\right|}^2.
\end{equation}
These products should be regularized. The standard way is to use the Schwinger proper time regularization. Namely we exploit the following relation (here $\gL$ is a regularizer, the dynamically generated scale):
\begin{equation}
\label{Schwinger}
\mathfrak{Reg} (\e^{ia}) \equiv {\left.\frac{\dd}{\dd s}\right|}_{s=0}\frac{\gL^s}{\gG(s)}\int_0^\infty \frac{\dd t}{t} t^s \e^{i t a} = \ln\left|\frac{a}{\gL}\right|.
\end{equation}
It follows that $Z_{pert}^{gauge} = {\left|\e^{-\sum_{l\neq m}^N \g_\eps(a_l-a_m)}\right|}^2$, where
\begin{equation}
\g_\eps(x) = \mathfrak{Reg}\left(\frac{\e^{ix}}{\e^{i\eps}-1} \right)  = {\left.\frac{\dd}{\dd s}\right|}_{s=0}\frac{\gL^s}{\gG(s)}\int_0^\infty \frac{\dd t}{t} t^s \frac{\e^{i t x}}{\e^{it\eps}-1} = \sum_{g=0}^\infty \eps^{g-1}\g_g(x)
\end{equation}
Operation $\mathfrak{Reg}$ is linear, which implies that $\g_\eps(x)$ satisfies the following ``first order'' difference equation:
\begin{equation}
\g_\eps(x+\eps) - \g_\eps(x) = \mathfrak{Reg}(\e^{ix}) = \ln \left|\frac{x}{\gL}\right|.
\end{equation}
This equation allows us to determine this function up to an additive constant. We have
\begin{equation}
\g_\eps(x) = \frac{x}{\eps}\left(\ln \left|\frac{x}{\gL}\right| -1\right) - \frac{1}{2}\ln\left|\frac{x}{\gL}\right| + \sum_{g=1}^\infty {\left(\frac{\eps}{x}\right)}^{2g-1}\frac{B_{2g}}{2g(2g-1)},
\end{equation}
where $B_{2g}$ are Bernoulli numbers. The four dimensional analog of this function is $\g_{\eps_1,\eps_2}(x)$ defined in the Appendix A of \cite{SWandRP}.

This expansion shows that $W_{pert}^{gauge}= O(\eps)$. At the same way one can show that 
\begin{equation}
\label{Wpert}
W_{pert}^{matter} = \sum_{f=1}^{N_F}\sum_{l=1}^N (a_l + m_f) \left( \ln\left|\frac{a_l+m_f}{\gL}\right|-1\right) - \sum_{f=1}^{N_{\tilde{F}}}\sum_{l=1}^N (a_l + \tilde{m}_f) \left( \ln\left|\frac{a_l+\tilde{m}_f}{\gL}\right|-1\right) + O(\eps).
\end{equation}
This form of the perturbative par of thr twisted superpotential implies the renormalization group equation \Ref{RGflow}.


\subsection{Non--perturbative part}

In order to compute vortex contribution to the twisted superpotential we use the finite dimensional model for the vortex moduli space. It the $\gO$--background the deformed BRST operator satisfies:
\begin{equation}
\label{Q_m^2}
{(\Q_m')}^2 = \G(\phi) + \F(m) + \L(\eps),
\end{equation}
where the last term is the Lorentz rotation on angle $\eps$. The finite dimensional version of equations \Ref{QtwistGauge}, \Ref{HypTwist} and \Ref{HypTwistBar} properly deformed is the following:
\begin{equation}
\label{finiteSUSY}
\begin{aligned}
\Q'_m C &= \psi_C & \Q'_m \psi_C &= i[\phi,C] - i\eps C \\
\Q'_m I &= \psi_I & \Q'_m \psi_I &= i\phi I - i I a - \frac{i\eps}{2}I \\
\Q'_m \chi &= B & \Q'_m B &= i[\phi,\chi] \\
\Q'_m \l &= \eta & \Q'_m \eta &= i[\phi,\l] \\
\Q'_m \nu &= w & \Q'_m w &= i\phi \nu + i m \nu - \frac{i\eps}{2}\nu.
\end{aligned}
\end{equation}
Here $\nu$ is fermion and $w$ is boson. The weight of the Lorentz rotations for $C$, $I$ and $w$ can be explained as follows. $w$ classifies solutions for the Dirac equation for $q = -x^-$ and therefore it transforms under Lorentz rotations as lower component of a spinor, that is by multiplying to $-\frac{i\eps}{2}$. The remark below \Ref{Vk} explains the weights for $C$ and $I$ (recall that $\bar{z}\mapsto \e^{-i\eps}\bar{z}$). 

Let us now construct the finite dimensional version for the full action $S_{full}$. Its value on a vortex solution is given by the first term and equals $2\pi i \tau k$. The rest is the Mathai--Quillen representative of the equivariant Euler class for the Dirac equation solutions bundle (Dirac bundle for short) over the vortex moduli space, which is defined by equations  \Ref{Vk}. The finite action is given by $S_{finite} = \Q_m' \Psi_{finite}$, where
\begin{equation}
\label{PsiFinite}
\Psi_{finite} = \chi (\mu -2r) + \psi_C [\l, C] + \bar{\psi}_C [\l, C^\dag] + \psi_I \l I - I^\dag [\l,\bar{\psi}_I] + \left(\bar{\nu}w + \bar{w} \nu\right).
\end{equation}
First term enforces the integral to localize on submanifold $\mu = 2r$ whereas the rest describes the action of $\UU(k)$. The last term is due to solutions for the Weyl equation in  vortex background.

Integrating out boson and fermion matrices we obtain the following expression:
\begin{equation}
\label{Wvort}
\e^{\frac{1}{\eps}W_{vort}(a,\tau,\eps)} = 1 + \sum_{k=1}^\infty \e^{2\pi i \tau} Z_k(a,\eps),
\end{equation}
where
\begin{equation}
\label{Z_k}
Z_k(a,\eps) = \frac{1}{k!}\frac{1}{\eps^k} \int \prod_{i=1}^k \frac{\dd \phi_i}{2\pi i} \e^{2ir \sum_{i=1}^k \phi_i}\prod_{i\neq j}^k\frac{\phi_i - \phi_j}{\phi_i - \phi_j - \eps} \prod_{i=1}^k \prod_{l=1}^N \frac{\phi_i +m_l- \eps/2}{\phi_i - a_l - \eps/2}. 
\end{equation}
More elegant way to get the same result is to apply once again the localization technique, now to the finite dimensional space $\V_k$. The integral for $Z_k(a,\eps)$ can be computed with the help of the Duistermaat--Heckman formula. The weights for the maximal torus action of $\UU(N)_G \times \UU(N)_F \times \UU(1)'_L$ can be obtained directly from \Ref{finiteSUSY}. Details of these computations  can be found in \cite{MooreYM2D,GeneralizedInst,DeformationInstanton,SWfromInst,MyThesis}. 

The integral \Ref{Z_k} can be computed by residues. To do this we move $\phi_i$ to the complex plane. Like in four dimensional case $\eps$ gains positive imaginary part: $\eps \mapsto \eps + i0$. The exponent in the integrand indicates that we have to close the contour of integration in the upper halfplane, since $r >0$. All poles of the integrand are in the upper halfplane. It seems that when $r < 0$ the integral \Ref{Z_k} vanishes, since we close contour in the lower halfplane. However, when $r < 0$ the roles of $\tilde{Q}$ and $Q$ are interchanged. Also the sign between the field strength and the matter fields in the Bogomol'ny equation \Ref{Bogomol} is changed, which implies that now $C$ should be identifies with $A_z$. Therefore we replace $\eps\mapsto - \eps$ in equation \Ref{finiteSUSY}. This implies that all residues now are in the lower halfplane and again captured by the contour integration.


\subsection{Residue handling}

Let us explain the manipulation with residues for the integral \Ref{Z_k}. In the $\Quat$--case the poles of similar integrals are enumerated by colored Young tableaux \cite{GeneralizedInst,SWfromInst}. Relevant formulae can be obtained in the context of Hilbert schemes of points on surfaces \cite{NakajimaJack,Nakajima,NakajimaLecture}. In our case the classification of residues can be obtained by similar approach, though more simple.

The residues are classified by $N$ icicles of heights $\vec{k} = \{k_1,\dots,k_N\}$. The total height of all icicles is $|\vec{k}| = k_1 + \dots + k_N = k$. The residues which correspond to given configuration of icicles are in points $\phi^\star_i = a_l + (\frac{1}{2} + i_l)\eps$, where $i_l = 0,1,\dots,k_l-1$. For such a configuration we can put forward the following formula (which can be proved by induction):
\begin{equation}
\label{Chern}
\sum_{i,j=1}^k \left(\e^{\phi_i^\star-\phi_j^\star} - \e^{\phi_i^\star-\phi_j^\star-\eps} \right) - \sum_{l=1}^N\sum_{i=1}^k \e^{\phi_i^\star - a_l - \eps/2} = - \sum_{l,m=1}^N \sum_{i_l=1}^{k_l}\e^{a_l-a_m + (k_l - k_m - i_l)\eps}.
\end{equation}
Now we apply to this identity Schwinger regularizing procedure $\mathfrak{Reg}$ defined in equation \Ref{Schwinger}, and transform the sum of exponents to the product of their arguments: $\sum_\a\e^{w_\a} \mapsto \prod_\a w_\a$. Number of possible ordering of $\phi_i$ is equal to $k!$ which compensates $\frac{1}{k!}$ factor in the integral \Ref{Z_k}. The exponents whose argumet depends only on $\eps$ and not on $a_l$ lead to the combinatorial  factor which is equal to the number of ways to distribute $k$ residues between $N$ icicles. 

The final expression for the integral \Ref{Z_k} is the following:
\begin{equation}
Z_k(a,\eps) = \sum_{\vec{k}:|\vec{k}| = k}\frac{1}{\vec{k}!\eps^{|\vec{k}|}}\e^{2ir\vec{k}\cdot \vec{a}}\frac{\prod_{f=1}^{N}\prod_{p=1}^N \prod_{i_p=1}^{k_p}(a_p + m_f + i_p \eps )}{\prod_{l\neq m}^N \prod_{i_l=1}^{k_l} \left(a_l - a_m + (k_l - k_m - i_l)\eps\right)},
\end{equation} 
where $\vec{k}\cdot \vec{a} = \sum_{i=1}^k \phi_i^\star = \sum_{l=1}^N \left(a_l k_l + \frac{k_l^2}{2}\eps\right)$ and $\vec{k}! = \prod_{l=1}^N (k_l !)$.

Formula \Ref{Wvort} allows us to relate this quantity to
\begin{equation}
\label{Wdevelop}
W_{vort}(a,\eps) = \sum_{g=0}^\infty \eps^g W_g^{vort}(a).
\end{equation}
The leading term of this series, that is, $W_0^{vort}(a)$, is the twisted superpotential introduced in \Ref{twistedSup}. 


\section{Concluding remarks}
\label{conclusion}


\subsection{About $\Compl$--case}

We have shown how the localization technique can be applied to study of two dimensional topological models. Morally speaking we have adapted the instanton counting story \cite{SWfromInst} to the two dimensional case. We have founded a lot of similarities. However some features of four dimensional theory can not be reproduced. It may be caused by low--dimensional effects (such as full breaking of the $\R$--symmetry group, instead of partial breaking in four dimensions), or may point to some pathological obstacles.

Recall that in the $\Quat$--case the rational factors for the integrand in formula \Ref{Z_k} can be obtained by applying the Schwinger regularization \Ref{Schwinger} to the equivariant Chern character of the Dirac bundle $\E$ (see for details \cite{SWfromInst,MyThesis,SPinSW}). 

If it were true in the $\Compl$--case, then the Chern character for the Dirac bundle for adjoint representation of the gauge group would be equal, roughly speaking, to the lefthand side of \Ref{Chern}. The moduli space data ($B$ and $I$) can be combined to linear map acting as follows:
\begin{equation}
\begin{CD}
V\otimes S_- \oplus W @>B\oplus I>> V\otimes L,
\end{CD}
\end{equation}
where $V = \Compl^k$, $W = \Compl^N$, $S_- = L = \Compl$. $V$ is acted on by $\UU(k)$, the space $W$ is acted on by $\UU(N)_G$. $S_-$ is the space of Dirac spinors with negative chirality and $L$ is a fiber of the determinant bundle. In the $\Quat$--case similar construction was a complex, but now since we have only one such map it is meaningless to call it so.

Consider an element of the product group torus $t \in\Tor_{\UU(k)} \times \Tor_{\UU(N)_G}\times \Tor_{\UU(1)_L}$. The equivariant Chern character can be computed as follows:
\begin{equation}
\Ch_t(\E) = \Tr_{\E} (t) = \Tr_W(t) + \Tr_V(t)\left(\Tr_{S_-}(t) - \Tr_L(t) \right) = \sum_{l=1}^N \e^{a_l} - (\e^{\eps}-1)\e^{-\eps/2} \sum_{i=1}^k \e^{\phi_i}.
\end{equation}
The equivariant index of the Dirac operator is given by the equivariant analog of the Atiyah--Singer theorem:
\begin{equation}
\Ind_t^{fund} \cD = \int_{\Compl} \Ch_t(\E) \Td_t(\Compl),
\end{equation}
where $\Td_t(\Compl) = \frac{\eps}{\e^{\eps}-1}$ is the equivariant Todd class for $\Compl$. The integral can be computed equivariantly, the moment map is given by $\mu = \eps{|z|}^2$. Then for the fundamental and adjoint representations of $\UU(N)_G$ we obtain
\begin{equation}
\begin{aligned}
\Ind_t^{fund} \cD &= \frac{\Ch_t(\E)}{\e^{\eps}-1} = \frac{\sum_{l=1}^N \e^{a_l}}{\e^{\eps}-1} - \e^{-\eps/2}\sum_{i=1}^k \e^{\phi_i}, \\
\Ind_t^{adj} \cD &= \int_{\Compl} \Ch_t(\E^\ast\otimes\E)\Td_t(\Compl) = \int_{\Compl} \Ch_t(\E^\ast)\Ch_t(\E)\Td_t(\Compl) \\
&= \frac{\sum_{l,m=1}^N \e^{a_l-a_m}}{\e^{\eps}-1} - \sum_{l=1}^N\sum_{i=1}^k \e^{\phi_i - a_l - \eps/2}  - \sum_{i,j=1}^k \e^{\phi_i-\phi_j}\left( 1 - \e^{-\eps}\right) + \sum_{l=1}^N\sum_{i=1}^k \e^{a_l-\phi_i-\eps/2}.
\end{aligned}
\end{equation}
We observe now that the first term (which contains an infinite number of summands) is converted by the regularization procedure \Ref{Schwinger} to the perturbative corrections to the partition function which are given by formulae \Ref{Zpert^matter} and \Ref{Zpert^gauge}. The rest of terms (but the last sum in the last line) are converted to the integrand of \Ref{Z_k}. 

The last term does not have its counterpart in the expression for the partition function. Also we can not reproduce in such a way the $\tilde{Q}$--contribution to the perturbative part of the partition function, whose leading term is given by the second sum in \Ref{Wpert}. One of possible explanation of such a behavior is that in the $\Compl$--case, in opposition to the $\Quat$--case, the moduli space \Ref{Vk} describes solutions for fields $A_\mu$ and $\tilde{q}$ and not for $A_\mu$ solo.

Another problem we meet is the absence of the analog of the Seiberg--Witten theory in two dimensions. Recall that in the $\Quat$--case the prepotential can be expanded as follows: $\F(a,\hbar) = \sum_{g=0}^\infty \hbar^{2g} \F_g(a)$, where $\hbar = \eps_1 = - \eps_2$. The leading term of this expansion, $\F_0(a)$, which is known as Seiberg--Witten prepotential, can be defined through the cycles of an algebraic curve \cite{SeibergWitten,SeibergWittenII}. This prescription appears naturally in the thermodynamical ($\eps_1,\eps_2\to0$) limit in the exact expressions analogous to \Ref{Z_k}, as it was shown by Nekrasov and Okounkov. See for details \cite{SWandRP,MyThesis,SPinSW}.

One can perform similar manipulations in $\Compl$--case as well. If the Nekrasov--Okounkov approach is valid, then in the limit $\eps\to0$ the sum \Ref{Wvort} is dominated by a single term with $k \sim \frac{1}{\eps}$. Introduce the vortex density normalized in such a way to remain finite in the thermodynamical limit: $\rho(x) = \eps \sum_{i=1}^k \d(x-\phi_i)$. The $k$tuple integral \Ref{Z_k} (and therefore the whole vortex partition function \Ref{Wvort}) can be approximated  by a path integral
\begin{equation}
\label{Zpath}
Z \sim Z_{\frac{1}{\eps}} \sim \int \D \rho \e^{-\frac{1}{\eps} \left( H[\rho] + O(\eps)\right)},
\end{equation}
where the Hamiltonian is given by 
\begin{equation}
H[\rho] = - \vpint\dd x \dd y \frac{\rho(x)\rho(y)}{x-y} - \sum_{f=1}^N \int\dd x \rho(x)\ln\left|x+m_f\right| + \sum_{l=1}^N\int \dd x \rho(x)\ln \left|x-a_l\right| - 2ir \int \dd x\rho(x) x. 
\end{equation}
The first term vanished for symmetry reason, and the line of arguments which lead to the $\Compl$--analog of the Seiberg--Witten theory fails, for it is based on the saddle point approximation for the path integral \Ref{Zpath}. It follows that the F--term contribution to the effective action \Ref{twistedSup}, which is defined by the twisted superpotential $W_0(a)$ can not be reproduced by a sort of cycle computation. 


\subsection{About $\Oct$--case}

Let us finally say a word about the eight dimensional theory. The partition function of the model in the $\gO$--background  will be localized onto the moduli space of the generalized instantons, which are defined as solutions of the equation in the second line of \Ref{genInst}. Presumably moduli space of such instantons has the finite dimensional realization which is given by  a straightforward generalization of \Ref{Vk} and the ADHM construction \cite{ADHM,SelfDualSolution,InstAndRec}. 

Then corresponding finite dimensional integrals for the partition function will have a similar form as corresponding integral which appear in Witten index computation performed in \cite{GeneralizedInst}. In a manner of speaking, modulo some technicalities, one can say that Witten index computation is ``dual'' to the instanton counting scheme. Indeed, in \cite{GeneralizedInst} the remaining integration in the counterpart of \Ref{Z_k} is taken over the maximal torus of the group of rigid gauge transformations, i.e. the gauge transformation at infinity, whereas while doing the generalized instanton counting the remaining integration is to be taken over the dual (in the sense of \cite{SelfDualSolution}) group. Recall that in the $\Compl$--case it is $\UU(k)$. This dual group is all what remains from the whole gauge group (which consists of all gauge transformations with fixed value at infinity) in the finite dimensional model for the moduli space. 

Apart from aesthetic wish to complete the $\Compl$--$\Quat$--$\Oct$ story, there is a purely pragmatic motivation to study eight dimensional model. Recall that the need to have the Fayet--Iliopoulos term in the $\Compl$--case forces us to focus on the $\UU(N)$ gauge group, which is the group of isometries of a complex vector space. The ADHM construction is known only for classical semi--simple groups, such as $\SU(N)$, $\SO(N)$ and $\Sp(N)$. This triad is directly related to the quaternion vector space isometries. It is plausible to believe that the moduli space of eight dimensional generalized instantons is related in some sense to isometries of an octonion vector space. The exceptional groups ($E$, $F$ and $G$ root systems) are closely related to such isometries \cite{Octonions}. Thus the finite dimensional construction should be valid for all semi--simple groups. We believe that the construction of $\Oct$--instantons can shed some light to the instanton counting in four dimensions, that is, to the construction of Seiberg--Witten prepotential.


\providecommand{\bysame}{\leavevmode\hbox to3em{\hrulefill}\thinspace}
\providecommand{\href}[2]{#2}

\end{document}